\documentclass[letterpaper,twocolumn,10pt]{article}
\usepackage{usenix-2020-09}
\usepackage{amssymb}
\usepackage{amsmath}
\usepackage{booktabs}
\usepackage{xcolor}
\usepackage{algorithm}
\usepackage{algpseudocode}
\algrenewcommand{\Return}{\textbf{return}~}
\usepackage{graphicx}
\usepackage{float}
\usepackage{placeins}
\usepackage{tikz}
\usepackage{enumitem}
\usepackage{pifont}
\usepackage{hyperref}
\usetikzlibrary{arrows.meta,positioning,shapes.geometric,fit,backgrounds,calc,decorations.pathreplacing}

\begin{document}

\date{}


\title{\Large \bf \textsc{KVMark}: Fragile Watermarking for KV-Cache Integrity\\in Latent Multi-Agent Systems}


\author{
{\rm Marjan Beheshti$^{1}$, Ronghua Xu$^{1}$}\\
$^{1}$Michigan Technological University\\
Houghton, MI, USA\\
mbehes2@mtu.edu, ronghuaxu@mtu.edu
}

\maketitle

\begin{abstract}
Latent multi-agent communication directly transfers Key-Value (KV) cache tensors between transformer-based agents, bypassing text serialization entirely.
While computationally efficient and reasoning accurate, this communication channel is opaque and vulnerable to man-in-the-middle tampering, a threat vector unaddressed by existing text-based or behavioral watermarks, which cannot operate on continuous internal representations.
To address this vulnerability, we propose \textsc{KVMark}, a training-free, zero-bit KV-cache watermarking framework to verify inter-agent handoff integrity.
Our framework introduces four core technical components:
(1)~position-dependent sign patterns that guarantee first-order statistical invisibility while preserving downstream attention survival;
(2)~per-head sign patterns with aligned detection to defeat head-averaging attacks;
(3)~a three-tier depth-aware strength schedule that equalizes detection sensitivity across transformer layers while maintaining invisibility against nine statistical profiling strategies;
and (4)~content-bound key derivation that cryptographically binds watermarks to input contexts, preventing replay attacks under Kerckhoffs's principle.
Empirical evaluation on a four-agent latent pipeline demonstrates $100\%$ detection across $1{,}500$ handoffs with a $0.0\%$ false positive rate ($\text{AUROC}=1.0$), $100\%$ layer-level tamper localization ($17/17$ attacked layers identified with $0/19$ false positives), and no statistically significant reasoning accuracy degradation across four benchmarks ($N{=}300$--$500$).
Furthermore, the framework enables multi-hop cascade verification to isolate compromised agents ($500/500$ detection at three hops), withstands adaptive white-box adversaries and the \textit{Out-of-Sight} DiffMean attack (up to $\alpha=8.0$), and generalizes seamlessly to Qwen3-8B without re-tuning ($1{,}500/1{,}500$ TPR).
\end{abstract}

\section{Introduction}
\label{sec:intro}

Large language models (LLMs) have rapidly evolved from standalone reasoners into autonomous agents capable of multi-step planning, tool invocation, and self-reflection~\cite{react,toolformer,autogen}. To solve tasks that exceed any single model's capacity, recent work composes multiple specialized agents into collaborative pipelines---\emph{multi-agent systems} (MAS)---where distinct roles such as planner, critic, refiner, and judge coordinate through structured inter-agent communication~\cite{latentmas,autogen,gsafeguard}.
Traditionally, multi-agent systems coordinate via natural language, which compresses rich internal states into discrete tokens but imposes a fundamental information bottleneck~\cite{latentmas,coconut}.
To maximize representational capacity, \emph{latent multi-agent systems} transmit internal KV caches $\{K^{(l)}, V^{(l)}\}_{l=1}^{L}$ directly between agents, thereby bypassing text serialization entirely~\cite{latentmas,interlat}. 
While this training-free paradigm achieves superior reasoning accuracy~\cite{latentmas}, transmitting high-dimensional continuous tensors simultaneously moves the communication channel beyond conventional inspection, filtering, or content moderation. This architectural shift introduces a critical security gap: an unscrutinized, continuous-valued attack surface lacking any native integrity-verification mechanism.


This vulnerability is concrete and empirically validated. Wang et al.~\cite{outofsight} show that inter-agent KV-cache handoffs constitute the most damaging attack vector in latent MAS: offline-extracted perturbations generalize broadly, while V-cache modifications remain invisible to output-quality checks. Crucially, current frameworks~\cite{latentmas,interlat} lack native integrity verification. While transport-layer mechanisms (e.g., HMACs) catch post-handoff substitution, they fail against compromised endpoints that generate and sign malicious KV in-band~\cite{latentagentslie}. Because downstream reasoning is causally dependent on upstream KV content---where removing a single specialist's hidden state causes nearly half of correct outputs to fail~\cite{latentagentslie}---unverified KV transfers pose a severe threat.
While watermarking is the natural candidate for this gap, existing methods fail in the latent setting: text watermarks~\cite{kgw,threebricks,robustwatermark,kuditipudi2024robust} require logit sampling during decoding, while agent watermarks~\cite{agentmark,agentwm,acthook,agentguide} target discrete action trajectories rather than continuous tensors. More fundamentally, watermarking a KV cache presents two unique challenges. First, to prevent keyless adversaries from profiling and removing the signal, the watermark must be \emph{statistically invisible}. Second, to detect compromised intermediate agents, an upstream watermark must \emph{survive downstream attention computation} so monitors can verify its presence at later pipeline stages. Achieving both properties simultaneously has not been demonstrated by prior work.

\begin{figure}[t]
\centerline{\includegraphics[width=0.48\textwidth]{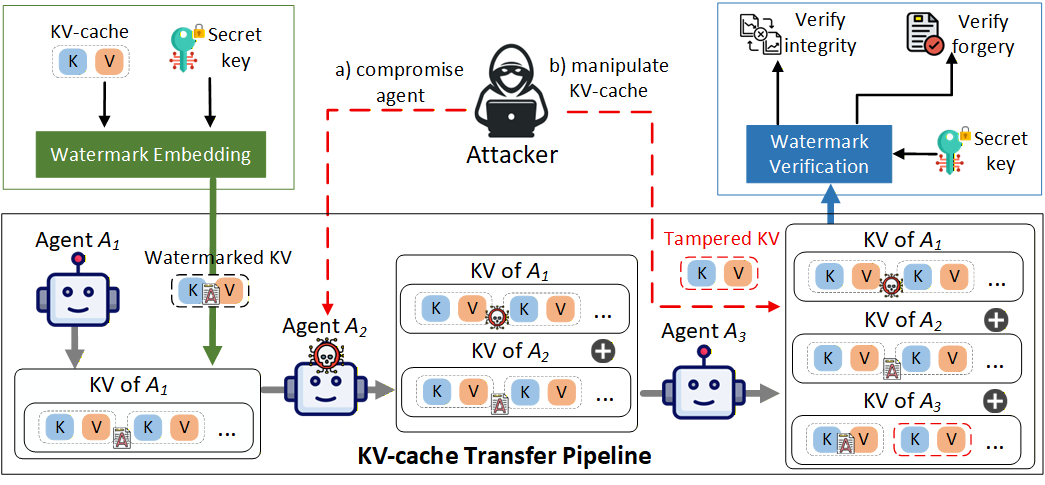}}
\caption{The High-level architecture of \textsc{KVMark}. \textbf{Top Left:} \emph{Watermark Embedding} applies an agent-specific key to embed a keyed perturbation into $K$ and $V$ tensors prior to transfer.
\iffalse a secret key to encode $K$ and $V$ tensors prior to transfer.\fi \textbf{Bottom:} The \emph{KV-cache Transfer Pipeline} maintains an append-only sequence of agent states ($A_1 \to A_2 \to A_3$), subject to adversarial agent compromise or KV manipulation. \textbf{Top Right:} The \emph{Watermark Verification} module evaluates incoming multi-hop caches against agent-specific keys to verify integrity and detect forged or tampered entries.}
\label{fig:system_concept}
\end{figure}

To address these challenges, this paper introduces \textsc{KVMark}, the first watermarking scheme specifically designed for KV-cache integrity verification in latent multi-agent systems.
As illustrated in Fig.~\ref{fig:system_concept}, our approach embeds fragile, zero-bit statistical watermarks into both the V-cache and K-cache at each inter-agent handoff using agent-specific keys, enabling a centralized verification module (operated by the pipeline defender) to confirm provenance, detect malicious tampering, and identify forgery attempts.
The \textsc{KVMark} framework's design addresses the core security requirements: position-dependent sign patterns guarantee first-order statistical invisibility (column-mean invariance) against keyless adversaries and per-head sign patterns with aligned detection mitigate head-averaging attacks, while depth-aware magnitude scaling preserves invisibility against higher-order statistical profiling (variance, kurtosis, range, absolute mean, and cross-cache correlation).
Furthermore, leveraging the append-only structure of autoregressive KV-cache handoffs across the pipeline, the embedded signal persists across downstream execution stages, enabling cascade verification to isolate compromised intermediate agents.


The watermark embedding and verification pipeline introduces three core innovations. First, to resolve the tension between attention survival and detectability, we propose position-dependent sign patterns that achieve both statistical invisibility and downstream propagation simultaneously.
Second, to address distinct depth-dependent decay profiles between V-cache and K-cache spaces, we introduce a three-tier V-cache strength schedule (suppress early, baseline middle, boost late) alongside uniform K-cache boosting, equalizing detection sensitivity across all transformer layers while maintaining full statistical invisibility.
Third, to mitigate replay attacks in multi-agent pipelines, we propose content-bound key derivation that cryptographically binds each watermark to both the agent's secret key and the specific input context, thereby satisfying Kerckhoffs's principle while preventing cross-query cache reuse.

Our main contributions are:
\begin{itemize}[nosep,leftmargin=1.5em]
    \item We propose \textsc{KVMark}, a training-free, zero-bit KV-cache watermarking framework designed to verify handoff integrity in latent multi-agent systems. Empirical evaluation across 1,500 independent handoff verifications demonstrates the scheme achieves 100\% TPR at 0.0\% FPR ($\text{AUROC}=1.0$) with no statistically significant accuracy degradation ($92.4\%$ vs.\ $92.0\%$ clean baseline on GSM8K).
    
    \item We characterize divergent V-cache and K-cache depth profiles and resolve them via a three-tier V-cache strength schedule ($0.1\times$ at layers 0--4, $1.0\times$ at layers 5--23, and $5.0\times$ at layers 24--35) and uniform K-cache boosting ($10\times$ across all layers). This calibration achieves 100\% layer-level attack localization ($17/17$ layers identified, $N=500$) and full statistical invisibility across nine detection strategies.
    
    \item We evaluate system resilience against a broad attack suite. Experimental results demonstrate 100\% detection against quantization, adaptive white-box attacks, the DiffMean attack~\cite{outofsight}, and a 99.8\% replay rejection rate ($N=500$). Finally, the framework enables cascade verification to detect compromised intermediate agents across multi-hop pipelines ($500/500$ at 3 hops).
\end{itemize}

\section{Background and Related Work}
\label{sec:related}

\subsection{Latent Multi-Agent Communication}
\label{sec:latent_mas}

In a standard transformer-based LLM, each layer $l \in \{1, \ldots, L\}$ maintains a key-value cache $\{K^{(l)}, V^{(l)}\}$ that stores the computed attention keys and values for all previously processed tokens, avoiding redundant recomputation during autoregressive generation~\cite{vaswani2017attention}. In text-based multi-agent systems, Agent~$A_i$ generates a text response that Agent~$A_{i+1}$ re-tokenizes and processes from its input embedding layer, discarding $A_i$'s internal representations entirely.

Latent multi-agent systems exploit a key observation: when all agents share the same model architecture, the KV cache from Agent~$A_i$ is dimensionally compatible with Agent~$A_{i+1}$'s attention layers. Rather than serializing to text, LatentMAS~\cite{latentmas} transfers the full cache $M_{A_i} = \{K^{(l)}, V^{(l)}\}_{l=1}^{L}$ directly, allowing the receiving agent to continue computation as if it had generated those representations itself. This \emph{latent handoff} preserves the sender's full hidden-state information across all $L$ layers and $H$ attention heads, eliminating the information bottleneck inherent in text-based communication. Empirically, latent communication yields consistent accuracy improvements over text-based pipelines on mathematical reasoning~\cite{gsm8k}, commonsense, and code-generation benchmarks, while requiring no additional training or architectural modification~\cite{latentmas}.

Concurrent work addresses complementary concerns in KV-based latent communication. Li et al.~\cite{obf} reduce the communication cost of full KV-cache relay by compressing the cache at inter-agent handoffs, retaining only 10--20\% of positions via attention-based selection with orthogonal residual injection. Asif et al.~\cite{lcguard} study the \emph{privacy} dimension: their LCGuard framework learns representation-level transformations on KV caches before transmission to limit the recoverability of agent-specific sensitive inputs, formalizing leakage through an adversarial reconstruction objective.
 Neither work addresses \emph{integrity} of the KV cache in transit, which is the focus of the present paper.

\subsection{Watermarking in Language Model Systems}
\label{sec:watermarking}

Watermarking techniques for language models fall into two broad categories distinguished by the abstraction level at which they operate: \emph{text-generation watermarks} that modify the token sampling process, and \emph{agent-level watermarks} that embed signals into behavioral trajectories.

\paragraph{Text-generation watermarks.}
KGW~\cite{kgw} partitions the vocabulary into green and red lists conditioned on a secret key and preceding tokens, biasing the sampling distribution toward green-list tokens for zero-bit detection via hypothesis testing. Three Bricks~\cite{threebricks} extends this framework with provably optimal detection boundaries and improved robustness guarantees. Subsequent work addresses specific trade-offs: Kuditipudi et al.~\cite{kuditipudi2024robust} achieve distortion-free watermarking by coupling the key schedule to a pseudorandom sequence that leaves the output distribution unchanged, while Zhao et al.~\cite{robustwatermark} provide provable robustness bounds against bounded text edits. A comprehensive empirical analysis by Piet et al.~\cite{satml_markmywords} reveals fundamental trade-offs between detectability, output quality, and robustness across these schemes, echoed by the theoretical treatment of Pang et al.~\cite{nofreeunch}. However, all text-generation watermarks operate at the \emph{logit level} during token sampling---a mechanism that is entirely absent at a latent KV-cache handoff, where no tokens are generated and the transferred object is a continuous-valued tensor.

\paragraph{Agent-level watermarks.}

A parallel line of work embeds provenance signals into agents' behavioral outputs rather than token logits. One group biases the sampling distribution over discrete agent decisions---planning steps~\cite{agentmark,agentguide}, tool-execution paths~\cite{agentwm}---to encode multi-bit or zero-bit identifiers, with detection via z-statistics or distribution-preserving sampling. A second group instead attributes trajectories after the fact, embedding signals into action-transition patterns or complementary trajectory encodings for position- and deletion-robust verification~\cite{trace,seqbehavioral}.
Across both groups, the watermarked object is a \emph{discrete action sequence}---tool calls, API invocations, step orderings---with no counterpart in a KV-cache transfer, where the communicated object is a set of continuous attention keys and values across layer dimension ($L$) and head dimension ($H$)

\paragraph{The KV-cache gap.}
Neither text-generation nor agent-level watermarks address the latent communication channel. More fundamentally, both families assume the watermarked object is either a token sequence or an action trace---discrete structures amenable to distribution biasing or sequence encoding. A KV cache is a continuous, high-dimensional tensor that undergoes nonlinear transformation through the receiving agent's attention layers, creating a survival challenge that has no analogue in existing watermarking literature. Understanding the specific attack mechanisms that exploit this unprotected channel is essential for designing an effective defense.

\subsection{Attacks on Latent Multi-Agent Systems}
\label{sec:attacks}

Security analysis of multi-agent systems has primarily focused on text-level threats: prompt injection~\cite{badagent}, adversarial role manipulation, and inter-agent deception through crafted natural-language messages~\cite{gsafeguard,massurvey}. These attacks exploit the interpretability of text-based communication---the same property that makes text amenable to filtering and moderation.

The emergence of latent communication introduces a fundamentally different attack surface. Wang et al.~\cite{outofsight} provide the first systematic study of attacks on KV-cache handoffs in LatentMAS, introducing the \emph{DiffMean} attack: given a small set of 
paired KV caches (one from a clean-correct execution, one 
from the corresponding text-level-attacked execution), the 
attacker computes the element-wise mean difference to extract 
a single perturbation direction that, when injected into a new 
clean cache at a handoff point, reproduces the attack's effect 
without reusing the adversarial text. This attack is notable for three properties that distinguish it from text-level threats. First, the perturbation operates in continuous latent space and produces no visible change in any text output at the injection point. Second, V-cache-only perturbations cause negligible accuracy change ($|\Delta\mathrm{acc}| < 0.01$), making the attack nearly undetectable by output-quality monitoring. Third, the extracted direction generalizes across inputs: a single perturbation vector retains 92\% of its degradation on held-out samples, eliminating the need for per-input attack crafting.

The study further reveals an asymmetry between cache components (under PCA extraction): K-cache perturbations (which modify attention routing) cause moderate accuracy drops (0.215), while combined KV perturbations cause the most severe degradation (0.398). This component-level characterization establishes that V-cache is the least disruptive component to modify, while K-cache and combined KV attacks represent the most damaging threat vectors. The paper's threat model---an adversary who can intercept and modify outgoing KV-cache handoffs before the downstream agent consumes them---defines precisely the man-in-the-middle scenario that motivates our integrity verification scheme.

Brito and Baquero~\cite{latentagentslie} study KV-cache integrity in a role-sequenced planner-specialist-coordinator architecture. Their system transfers both a short visible text commitment and the full KV-cache hidden state separately to the coordinator; because these are independent channels, an attacker can corrupt the hidden-state channel while leaving the visible text intact, rendering text-level inspection ineffective. Their HMAC-SHA256 transport manifest achieves perfect detection for post-handoff substitution (774/774 honest payloads accepted, 295/295 tampered rejected), but their evaluation explicitly identifies a protection gap: compromised endpoints that sign malicious KV in-band produce valid signatures, and their magnitude quarantine is defeated by adaptive norm-matched attacks (0/65 detected under white-box optimization). Our watermark-based approach targets precisely this gap: the keyed statistical signal is embedded in the KV-cache values themselves, so even a compromised endpoint cannot produce a validly watermarked cache without the secret key.

\section{Problem Formulation}
\label{sec:problem}
We consider a sequential latent multi-agent system comprising $N$ agents $\{A_1, \ldots, A_N\}$ that share the same transformer architecture with $L$ layers and $H$ attention heads per layer. At each handoff, Agent~$A_i$ transfers its KV cache $M_{A_i} = \{K^{(l)}, V^{(l)}\}_{l=1}^{L}$ to Agent~$A_{i+1}$, where $K^{(l)}, V^{(l)} \in \mathbb{R}^{B \times H \times S \times d}$ denote the key and value tensors at layer $l$, with batch size $B$, sequence length $S$, and head dimension $d$.
{The pipeline operator holds a single master secret $sk$, confidential to the defender.
 Each agent $A_i$'s watermarking key is derived as $sk_i = \texttt{SHA-256}(sk \,\|\, \texttt{role}_i)$, from which the V-cache and K-cache watermarking keys follow: $k_V = sk_i$ and $k_K = sk_i \| \texttt{``\_K''}$ (Section~\ref{sec:embedding}). Deriving $sk_i$ from a single master secret rather than issuing $N$ independent secrets simplifies key management: the defender need only protect one value, and per-agent keys can be recomputed on demand from $sk$ and the (public) role name. Each agent receives only its own derived key $sk_i$ through a secure initialization channel; agents cannot derive other agents' keys.}
Table~\ref{tab:notation} summarizes the notation used throughout the paper.
\begin{table}[t]
\centering
\small
\caption{Notation summary.}
\label{tab:notation}
\setlength{\tabcolsep}{4pt}
\begin{tabular}{@{}ll@{}}
\toprule
\textbf{Symbol} & \textbf{Definition} \\
\midrule
$A_1, \ldots, A_N$ & Sequential agents in the pipeline \\
$L, H, d, S$ & Layers, heads, head dimension, seq.\ length \\
$K^{(l)}, V^{(l)}$ & Key and value tensors at layer $l$ \\
$M_{A_i},\; \tilde{M}_{A_i}$ & Original / watermarked KV cache \\
$\hat{M}$ & Received cache at detection (possibly tampered) \\
\midrule
$sk$ & Master secret (defender-held) \\
$sk_i,\; sk_i'$ & Per-agent key; content-bound variant \\
$k_V,\; k_K$ & V-cache and K-cache derived keys \\
\midrule
$\mathcal{D}_V^{(l)},\; \mathcal{D}_K^{(l)}$ & Watermark dimensions at layer $l$ ($n_d$ each) \\
$\sigma^{(l,h)},\; \sigma_K^{(l,h)}$ & Per-head position-dependent sign patterns \\
$\bar{V}^{(l)},\; \bar{K}^{(l)}$ & Aligned head-averaged cache values (Eq.~\ref{eq:aligned_avg}) \\
$e_V^{(l)},\; e_K^{(l)}$ & Expected signal after alignment ($= \alpha^{(l)}$) \\
$\alpha_V^{(l)},\; \alpha_K^{(l)}$ & Per-layer embedding strengths \\
$\alpha_{V,0},\; \beta_V,\; \gamma_V$ & V-cache base, boost ($5{\times}$), suppress ($0.1{\times}$) \\
$\alpha_{K,0},\; \beta_K$ & K-cache base, boost ($10{\times}$) \\
\midrule
$\mathcal{S}_V,\; \mathcal{S}_K,\; \mathcal{S}$ & V, K, and combined detection scores \\
$\tau$ & Detection threshold \\
\bottomrule
\end{tabular}
\end{table}

\subsection{Threat Model}
\label{sec:threat_model}

\paragraph{Assumptions.}
Adhering to Kerckhoffs's principle, we assume the adversary possesses comprehensive knowledge of the watermarking scheme---including the embedding algorithm, detection procedure, and the correlation-based detection formula---while the master secret {$sk$} remains strictly confidential. The adversary has access to watermarked KV caches (by intercepting handoffs) but cannot observe the key derivation process{, so knowing an agent's public role name does not reveal $sk_i$}. We target man-in-the-middle adversaries who intercept and modify caches \emph{between} agents. A compromised agent that uses its own key $sk_i$ to re-watermark corrupted output is partially addressed through cascade verification (Section~\ref{sec:cascade}). A fully compromised defender whose master secret $sk$ is leaked is outside scope, as this requires trust establishment mechanisms beyond watermarking.

\paragraph{Adversary models.}
We evaluate three distinct adversarial objectives:
\begin{enumerate}[nosep,leftmargin=1.5em]
    \item \textbf{Tampering attackers} aim to modify the KV cache in transit to steer downstream agent behavior while evading watermark detection. This includes additive perturbations (Gaussian noise, steering vectors), targeted attacks (DiffMean~\cite{outofsight}), and cache quantization (int8, fp16). The attacker may target V-cache only, K-cache only, or both components, and may restrict modifications to specific transformer layers.
    \item \textbf{Forgery attackers} attempt to produce a KV cache that passes watermark verification without possessing the secret key. Strategies include {wrong-key forgery}, statistical estimation of the watermark pattern from observed caches (average-and-subtract), and blind dimension zeroing.
    \item \textbf{Replay attackers} intercept a legitimately watermarked cache from a prior query and re-inject it as the response to a different query, exploiting the fact that the watermark itself is valid even though the content is contextually incorrect.
\end{enumerate}

\paragraph{Defender objectives.}
The defender (pipeline operator) holds the master secret $sk$ and maintains white-box access to all agents, deriving each $sk_i$ on demand as needed.
The verification module performs \emph{zero-bit} detection: it verifies handoff authenticity (``did this cache come from {agent $A_i$}?'') rather than embedding multi-bit identity payloads. In a fixed pipeline with known agent roles, zero-bit detection is sufficient because the topology already determines sender identity at each handoff. Upon receiving a KV cache at each handoff point, the module fulfills three objectives:
\begin{enumerate}[nosep,leftmargin=1.5em]
    \item \textbf{Integrity verification}: determine whether the received cache was produced by the expected sender and has not been modified in transit.
    \item \textbf{Tamper localization}: if tampering is detected, identify (a) which handoff edge was compromised, and (b) which transformer layers were affected.
    \item \textbf{Invisibility}: the watermark must not be detectable by a keyless adversary through statistical analysis of the cache values, ensuring that knowledge of the watermarking scheme alone is insufficient to locate or remove the embedded signal.
\end{enumerate}
\section{Method}
\label{sec:method}


\subsection{Overview}
\label{sec:overview}

\begin{figure*}[t]
\centering
\includegraphics[width=0.9\textwidth]{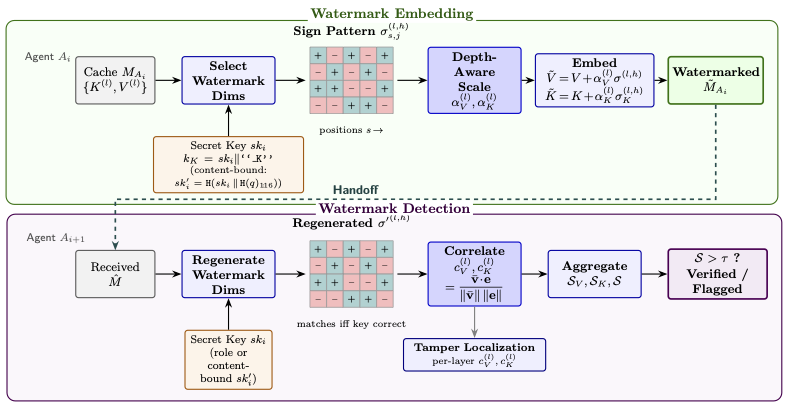}
\caption{Watermark embedding (top) and detection (bottom) in \textsc{KVMark}. 
The sending agent selects watermark dimensions via 
SHA-256, generates a per-head position-dependent sign pattern 
$\sigma^{(l,h)}_{s,j}$, applies depth-aware boosting, and 
embeds additive perturbations into both V-cache and 
K-cache. At the handoff, the verification module regenerates 
the expected dimensions and signs from the shared key, 
correlates with the received cache, and aggregates 
per-layer scores into $\mathcal{S}_V$, $\mathcal{S}_K$, and $\mathcal{S}$. 
Per-layer scores enable tamper localization.}
\label{fig:pipeline}
\end{figure*}

The integrity verification scheme of \textsc{KVMark} operates at each inter-agent handoff in the latent MAS pipeline. The protocol consists of two operations: \textbf{embedding}, performed by the sending agent after generation, and \textbf{detection}, performed by the verification module (operated by the pipeline defender) before processing. Figure~\ref{fig:pipeline} illustrates the complete pipeline.

Given a sequential pipeline $A_1 \to A_2 \to \cdots \to A_N$, each handoff $A_i \to A_{i+1}$ proceeds as follows:
\begin{enumerate}[nosep,leftmargin=1.5em]
    \item Agent $A_i$ completes its generation, producing KV cache $M_{A_i}$.
   \item The embedding module computes a keyed watermark signal from the secret key $sk_i$ and embeds it into both the V-cache and K-cache of $M_{A_i}$, producing $\tilde{M}_{A_i}$.
    \item The watermarked cache $\tilde{M}_{A_i}$ is transmitted to Agent $A_{i+1}$.
\item  Before processing, the verification module (operated by the pipeline defender) verifies $\tilde{M}_{A_i}$ using key $sk_i$, producing detection scores $\mathcal{S}_V$, $\mathcal{S}_K$, and their combined score $\mathcal{S}$. If $\mathcal{S}$ falls below a threshold $\tau$, the handoff is flagged as potentially tampered.
    \item If verification passes, Agent $A_{i+1}$ proceeds with its own generation using the received cache.
\end{enumerate}

The watermark is \emph{fragile by design}: any modification to the KV cache that is strong enough to affect downstream reasoning also disrupts the embedded signal, ensuring there is no ``silent tampering'' window where an attacker causes harm without detection.
The key $sk_i$ is derived from the agent role and, optionally, the input context (Section~\ref{sec:content_bound}), providing both sender authentication and replay resistance.

\subsection{Watermark Embedding}
\label{sec:embedding}

The embedding design must satisfy two requirements simultaneously. First, the watermark must be \emph{statistically invisible}: a keyless adversary who intercepts the cache in transit must not be able to identify which dimensions carry the signal, since an attacker who locates the watermark can surgically remove it before the receiving agent verifies. Second, for cascade verification across pipeline stages (Section~\ref{sec:cascade}), the watermark must \emph{propagate through attention computation}: when Agent~$A_{i+1}$ processes the received cache, each attention head computes $\mathrm{out} = \mathrm{softmax}(QK^\top / \sqrt{d})\, V$, a convex combination across positions ($\sum_i \alpha_i = 1$). A uniform perturbation $+\delta$ along a dimension therefore propagates exactly ($\mathrm{out} = \mathrm{original} + \delta$), whereas a perturbation at a specific position is diluted by its attention weight. We note that the KV cache is append-only: the receiving agent's generation appends new positions but does not modify existing ones. Per-handoff verification, which checks original positions before processing, is therefore agnostic to embedding strategy; the propagation requirement applies specifically to cascade verification of newly generated positions (Section~\ref{sec:cascade}). This propagation argument applies to V-cache; K-cache values determine attention weights rather than being convexly combined, so K-cache watermarks do not propagate to newly generated positions. However, since the cache is append-only, cascade verification of K-cache checks original positions directly.

These two requirements create a design tension: a uniform perturbation along selected dimensions propagates through attention but is trivially detectable via column-mean analysis. We resolve this tension through a combination of keyed dimension selection (which hides \emph{which} dimensions carry the signal) and position-dependent sign patterns (which prevent column-mean analysis from \emph{detecting} the signal even at the correct dimensions).

\paragraph{Keyed Dimension Selection.}
For each transformer layer $l$, we select subsets of $n_d$ watermark dimensions $\mathcal{D}_V^{(l)}, \mathcal{D}_K^{(l)} \subset \{1, \ldots, d\}$ using a keyed hash function:
\begin{equation}
    \mathcal{D}_V^{(l)} = \texttt{Select}\big(\texttt{SHA-256}(sk_i \,\|\, \texttt{``dims''} \,\|\, l),\; d,\; n_d\big),
    \label{eq:dim_select}
\end{equation}
where $\|$ denotes string concatenation, $\texttt{``dims''}$ is a fixed domain separator that ensures dimension selection and sign generation (Eq.~\ref{eq:sign_pattern}) produce independent hash outputs, and $\texttt{Select}(\cdot)$ deterministically maps the hash output to $n_d$ distinct indices. For K-cache, we define $k_K = sk_i \| \texttt{``\_K''}$ and compute dimensions independently:
\begin{equation}
    \mathcal{D}_K^{(l)} = \texttt{Select}\big(\texttt{SHA-256}(k_K \,\|\, \texttt{``dims''} \,\|\, l),\; d,\; n_d\big),
    \label{eq:dim_select_k}
\end{equation}
where $\texttt{``\_K''}$ is a domain separator ensuring V-cache and K-cache select independent dimensions from the same key. Without knowledge of $sk_i$ (and hence $sk$), an adversary cannot determine which dimensions carry the watermark.

\paragraph{Position-dependent Sign Pattern.}
Even with keyed dimension selection, a constant 
bias is detectable: an attacker computes $\mu_j = \frac{1}{S}\sum_{s=1}^{S} V_{s,j}$ for each dimension $j$; under a constant bias, the $n_d$ watermarked dimensions show $|\mu_j| \approx \delta$ while others show $|\mu_j| \approx 0$, immediately revealing the watermark. To defeat this, we introduce a \emph{position-dependent sign pattern} $\sigma \in \{-1, +1\}^{S \times n_d}$, a pseudorandom matrix of $\pm 1$ values keyed by the layer index and secret key. The pseudorandom signs yield approximately balanced columns ($\mathbb{E}[\bar{\sigma}_j] \approx 0$), making watermarked dimensions statistically indistinguishable from non-watermarked ones under column-mean analysis.

\paragraph{Per-head Sign Pattern.}
A second attack vector arises from the multi-head structure of the KV cache: an adversary who averages across all $H$ attention heads can suppress natural per-head variation while preserving a uniform watermark signal. To defeat this, we derive an \emph{independent} sign pattern for each head $h \in \{1, \ldots, H\}$ by including the head index in the PRG seed:
\begin{equation}
    \sigma^{(l,h)} = \texttt{PRG}\big(\texttt{SHA-256}(sk_i \,\|\, \texttt{``possig''} \,\|\, l \,\|\, S \,\|\, h),\; S,\; n_d\big),
    \label{eq:sign_pattern}
\end{equation}
\begin{equation}
    \sigma_K^{(l,h)} = \texttt{PRG}\big(\texttt{SHA-256}(k_K \,\|\, \texttt{``possig''} \,\|\, l \,\|\, S \,\|\, h),\; S,\; n_d\big),
    \label{eq:sign_pattern_k}
\end{equation}
where $\texttt{PRG}(\cdot)$ denotes a seeded generator producing $\pm 1$ values, $\texttt{``possig''}$ is a domain separator distinguishing sign generation from dimension selection (Eq.~\ref{eq:dim_select}), and $h \in \{1, \ldots, H\}$ is the attention head index. Since each head's sign pattern is independent, head-averaging cancels the watermark: $\frac{1}{H}\sum_h \alpha \cdot \sigma^{(l,h)}_{s,j} \approx 0$. The keyholder recovers the signal via \emph{aligned detection} (Section~\ref{sec:detection}), which multiplies each head by its known signs before averaging.

The watermark is embedded by additive perturbations to \emph{both} the V-cache and K-cache at each layer.
For each layer $l$, head $h \in \{1, \ldots, H\}$, position $s \in \{1, \ldots, S\}$, and watermark dimension $j$:
\begin{align}
    \tilde{V}^{(l)}_{h,s,j} &= V^{(l)}_{h,s,j} + \alpha_V^{(l)} \cdot \sigma^{(l,h)}_{s,j}, \quad j \in \mathcal{D}_V^{(l)}, \label{eq:embed_v} \\
    \tilde{K}^{(l)}_{h,s,j} &= K^{(l)}_{h,s,j} + \alpha_K^{(l)} \cdot \sigma_{K}^{(l,h)}{}_{s,j}, \quad j \in \mathcal{D}_K^{(l)}, \label{eq:embed_k}
\end{align}
where $\alpha_V^{(l)}$ and $\alpha_K^{(l)}$ are per-layer embedding strengths (Section~\ref{sec:depth_aware}), and all dimension sets and sign patterns follow Eqs.~\ref{eq:dim_select}--\ref{eq:sign_pattern} with key $sk_i$ for V-cache and $k_K = sk_i \| \texttt{``\_K''}$ for K-cache.
Using independent keys ensures that the V and K watermarks occupy different dimensions with overwhelming probability, avoiding interference.
The perturbation magnitude is identical across all batch elements and attention heads; however, the per-head sign patterns ensure that no cross-head averaging strategy can recover the watermark signal without the key (Section~\ref{sec:security}). Dimensions outside $\mathcal{D}_V^{(l)}$ and $\mathcal{D}_K^{(l)}$ remain unchanged.

\begin{algorithm}[t]
\caption{Watermark Embedding (\textsc{Embed})}
\label{alg:embed}
\begin{algorithmic}[1]
\Require Cache $M = \{K^{(l)}, V^{(l)}\}_{l=1}^{L}$ with seq.\ length $S$, key $sk_i$, strengths $\alpha_{V,0}$, $\alpha_{K,0}$, dims $n_d$
\Ensure Watermarked cache $\tilde{M}$
\State $k_K \gets sk_i \| \texttt{``\_K''}$
\For{$l = 1$ to $L$}
    \State $\mathcal{D}_V^{(l)} \gets \texttt{Select}(\texttt{SHA-256}(sk_i \| \texttt{``dims''} \| l), d, n_d)$
    \State $\mathcal{D}_K^{(l)} \gets \texttt{Select}(\texttt{SHA-256}(k_K \| \texttt{``dims''} \| l), d, n_d)$
    \State $\alpha_V^{(l)} \gets \textsc{Scale}_V(l, \alpha_{V,0})$ \Comment{Eq.~\ref{eq:boost_v}}
    \State $\alpha_K^{(l)} \gets \textsc{Scale}_K(l, \alpha_{K,0})$ \Comment{Eq.~\ref{eq:boost_k}}
    \For{$h = 1$ to $H$} \Comment{Per-head sign patterns}
    \State $\mathit{seed} \gets \texttt{SHA-256}(sk_i \| \texttt{``possig''} \| l \| S \| h)$
    \State $\sigma^{(l,h)} \gets \texttt{PRG}(\mathit{seed},\; S,\; n_d)$
    \State $\mathit{seed}_K \gets \texttt{SHA-256}(k_K \| \texttt{``possig''} \| l \| S \| h)$
    \State $\sigma_K^{(l,h)} \gets \texttt{PRG}(\mathit{seed}_K,\; S,\; n_d)$
    \For{$s = 1$ to $S$}
        \For{each $j \in \mathcal{D}_V^{(l)}$}
            $V^{(l)}_{h,s,j} \mathrel{+}= \alpha_V^{(l)} \cdot \sigma^{(l,h)}_{s,j}$
        \EndFor
        \For{each $j \in \mathcal{D}_K^{(l)}$}
            $K^{(l)}_{h,s,j} \mathrel{+}= \alpha_K^{(l)} \cdot \sigma_K^{(l,h)}{}_{s,j}$
        \EndFor
    \EndFor
    \EndFor
\EndFor
\State $\mathbf{return}\ \tilde{M} = \{\tilde{K}^{(l)}, \tilde{V}^{(l)}\}_{l=1}^{L}$
\end{algorithmic}
\end{algorithm}
\subsection{Watermark Detection}
\label{sec:detection}

Given a received KV cache and the secret key $sk_i$, the detection module reconstructs the expected watermark signals for both V-cache and K-cache and measures their correlation with the actual values. For each layer $l$, the detector recomputes the watermark dimensions $\mathcal{D}_V^{(l)}$, $\mathcal{D}_K^{(l)}$ and per-head sign patterns $\sigma^{(l,h)}$, $\sigma_K^{(l,h)}$ using the same key-derived hash functions as the embedding module (Eqs.~\ref{eq:dim_select}--\ref{eq:sign_pattern}), with key $sk_i$ for V-cache and $k_K = sk_i \| \texttt{``\_K''}$ for K-cache.

Detection employs \emph{aligned averaging}: rather than directly averaging cache values across heads, the detector first multiplies each head's received cache $\hat{V}^{(l)}_h$ (from $\hat{M}$) by its known sign pattern $\sigma^{(l,h)}$, which aligns all heads to a constant $+\alpha$ signal (since $\sigma^2 = 1$), then averages across heads:
\begin{equation}
    \bar{V}^{(l)}_{s,j} = \frac{1}{H}\sum_{h=1}^{H} \hat{V}^{(l)}_{h,s,j} \cdot \sigma^{(l,h)}_{s,j},
    \label{eq:aligned_avg}
\end{equation}
yielding the same noise-reduction benefit as standard head-averaging while preserving the watermark signal. The expected signal after alignment is constant: $e^{(l)}_{s,j} = \alpha_V^{(l)}$ for all $(s,j)$.

The per-layer detection scores are normalized dot products between the expected signals and the aligned cache values at the watermark dimensions:
\begin{align}
    c_V^{(l)} &= \frac{\sum_{s=1}^{S'} \sum_{j \in \mathcal{D}_V^{(l)}} \bar{V}^{(l)}_{s,j} \cdot \alpha_V^{(l)}}{\left\| \bar{\mathbf{v}}^{(l)} \right\| \cdot \left\| \mathbf{e}_V^{(l)} \right\|}, \label{eq:corr_v} \\
    c_K^{(l)} &= \frac{\sum_{s=1}^{S'} \sum_{j \in \mathcal{D}_K^{(l)}} \bar{K}^{(l)}_{s,j} \cdot \alpha_K^{(l)}}{\left\| \bar{\mathbf{k}}^{(l)} \right\| \cdot \left\| \mathbf{e}_K^{(l)} \right\|}, \label{eq:corr_k}
\end{align}
where $S'$ denotes the detection position range: $S' = S$ for standard per-handoff verification, and $S' = S_{\text{embed}}$ (the original embedding length) for cascade verification across pipeline stages (Section~\ref{sec:cascade}). $\bar{V}^{(l)}_{s,j}$ and $\bar{K}^{(l)}_{s,j}$ are the aligned head-averaged values (Eq.~\ref{eq:aligned_avg}), and $\|\cdot\|$ is the $\ell_2$ norm.

The aggregate detection score combines V-cache and K-cache correlations:
\begin{equation}
    \mathcal{S}_V = \frac{1}{L} \sum_{l=1}^{L} c_V^{(l)}, \quad
    \mathcal{S}_K = \frac{1}{L} \sum_{l=1}^{L} c_K^{(l)}, \quad
    \mathcal{S} = \frac{\mathcal{S}_V + \mathcal{S}_K}{2}.
    \label{eq:score}
\end{equation}
A handoff is classified as \emph{verified} if $\mathcal{S} > \tau$, where $\tau$ is the detection threshold. We set $\tau = 0.01$ in all experiments. The empirical gap between correct-key scores (min {$0.291$}) and wrong-key scores (max {$0.0070$}) means any $\tau \in ({0.0070, 0.291})$ achieves 0\% FPR and 100\% TPR; our $\tau = 0.01$ falls within this gap, yielding {0.0\% FPR ($0$ false positives across $1{,}500$ handoffs)}.
The individual scores $\mathcal{S}_V$ and $\mathcal{S}_K$ provide diagnostic information: a targeted V-only attack reduces $\mathcal{S}_V$ while leaving $\mathcal{S}_K$ intact, and vice versa. The per-layer scores $\{c_V^{(l)}, c_K^{(l)}\}_{l=1}^{L}$ are additionally retained for layer-level tamper localization (Section~\ref{sec:depth_aware}).

\begin{algorithm}[t]
\caption{Watermark Detection (\textsc{Detect})}
\label{alg:detect}
\begin{algorithmic}[1]
\Require Cache $\hat{M}$, key $sk_i$, strengths $\alpha_{V,0}$, $\alpha_{K,0}$, dims $n_d$, threshold $\tau$, position limit $S'$
\Ensure $(\mathcal{S},\; \text{verified},\; \{c_V^{(l)}, c_K^{(l)}\}_{l=1}^{L})$
\State $k_K \gets sk_i \| \texttt{``\_K''}$
\For{$l = 1$ to $L$}
    \State $\mathcal{D}_V^{(l)} \gets \texttt{Select}(\texttt{SHA-256}(sk_i \| \texttt{``dims''} \| l), d, n_d)$
    \State $\mathcal{D}_K^{(l)} \gets \texttt{Select}(\texttt{SHA-256}(k_K \| \texttt{``dims''} \| l), d, n_d)$
    \State $\alpha_V^{(l)} \gets \textsc{Scale}_V(l, \alpha_{V,0})$; $\alpha_K^{(l)} \gets \textsc{Scale}_K(l, \alpha_{K,0})$
    \For{$h = 1$ to $H$} \Comment{Per-head sign patterns}
    \State $\mathit{seed} \gets \texttt{SHA-256}(sk_i \| \texttt{``possig''} \| l \| S' \| h)$
    \State $\sigma^{(l,h)} \gets \texttt{PRG}(\mathit{seed},\; S',\; n_d)$
    \State $\mathit{seed}_K \gets \texttt{SHA-256}(k_K \| \texttt{``possig''} \| l \| S' \| h)$
    \State $\sigma_K^{(l,h)} \gets \texttt{PRG}(\mathit{seed}_K,\; S',\; n_d)$
    \EndFor
    \State $\bar{\mathbf{v}}^{(l)} \gets \frac{1}{H}\sum_{h} \hat{V}^{(l)}_{h}[\mathcal{D}_V^{(l)}] \odot \sigma^{(l,h)}$ \Comment{Eq.~\ref{eq:aligned_avg}}
    \State $\bar{\mathbf{k}}^{(l)} \gets \frac{1}{H}\sum_{h} \hat{K}^{(l)}_{h}[\mathcal{D}_K^{(l)}] \odot \sigma_K^{(l,h)}$
    \State $c_V^{(l)} \gets \cos(\bar{\mathbf{v}}^{(l)},\; \alpha_V^{(l)} \cdot \mathbf{1})$ \Comment{$\mathbf{1} \in \mathbb{R}^{S' \times n_d}$; Eq.~\ref{eq:corr_v}}
    \State $c_K^{(l)} \gets \cos(\bar{\mathbf{k}}^{(l)},\; \alpha_K^{(l)} \cdot \mathbf{1})$ \Comment{Eq.~\ref{eq:corr_k}; signal $= +\alpha$}
\EndFor
\State $\mathcal{S} \gets \frac{1}{2}\!\left(\frac{1}{L}\sum_l c_V^{(l)} + \frac{1}{L}\sum_l c_K^{(l)}\right)$
\State $\mathbf{return}\ (\mathcal{S},\; \mathcal{S} > \tau,\; \{c_V^{(l)}, c_K^{(l)}\}_{l=1}^{L})$
\end{algorithmic}
\end{algorithm}

\subsection{Depth-Aware Strength Scheduling}
\label{sec:depth_aware}
A fixed embedding strength leads to depth-dependent detection sensitivity decay.
The V-cache and K-cache exhibit distinct behavior: V-cache magnitude grows with transformer depth, causing the normalized correlation (Eq.~\ref{eq:corr_v}) to diminish at late layers; K-cache sensitivity is uniformly low across all layers. Appendix~\ref{sec:appendix_decay} provides full per-layer magnitude and correlation profiles across all 36 layers.
To equalize detection sensitivity, we introduce independent strength schedules for V and K, defining the per-layer strength functions $\textsc{Scale}_V$ and $\textsc{Scale}_K$ 
(Algorithms~\ref{alg:embed}--\ref{alg:detect}) as:
\begin{align}
    \alpha_V^{(l)} &= \begin{cases} \alpha_{V,0} \cdot \gamma_V & \text{if } l < l_{\text{suppress}}^V \\ \alpha_{V,0} & \text{if } l_{\text{suppress}}^V \leq l < l_{\text{boost}}^V \\ \alpha_{V,0} \cdot \beta_V & \text{if } l \geq l_{\text{boost}}^V \end{cases}, \label{eq:boost_v} \\
    \alpha_K^{(l)} &= \alpha_{K,0} \cdot \beta_K \quad \forall\, l, \label{eq:boost_k}
\end{align}
where $l_{\text{suppress}}^V = 5$ and $\gamma_V = 0.1$ suppress the V-cache at early layers (where the natural cache magnitude is comparable to the perturbation, making the watermark detectable by higher-order statistics such as variance and absolute mean), $l_{\text{boost}}^V = 24$ and $\beta_V = 5$ boost the V-cache at late layers (where the decay is most severe), while $\beta_K = 10$ boosts the K-cache uniformly across all layers (since K requires amplification everywhere). In all experiments, $\alpha_{V,0} = \alpha_{K,0} = 0.03$. These values are set empirically from the per-layer cache magnitude profile: suppress covers layers where V-cache std is comparable to the perturbation ($<0.15$), boost begins where uniform correlation drops below $0.08$, and $\beta_V = 5$ is the highest multiplier that maintains full statistical invisibility ($\leq 3.0/16$; see Appendix~\ref{sec:appendix_decay} for detailed per-layer analysis).

The three-tier schedule ensures that the perturbation magnitude remains a small fraction of each layer's natural signal at every depth: position-dependent sign patterns defeat first-order statistics (column-mean), while magnitude scaling defeats higher-order statistics (variance, kurtosis, range, absolute mean, cross-cache correlation) (Section~\ref{sec:security}). Additionally, per-head sign patterns (Section~\ref{sec:embedding}) defeat head-averaging attacks.
The per-layer scores $c_V^{(l)}$ and $c_K^{(l)}$ (Eqs.~\ref{eq:corr_v}--\ref{eq:corr_k}) directly enable tamper localization: a layer affected by tampering shows a statistically significant drop in its correlation score relative to unaffected layers.

\subsection{Content-Bound Key Derivation}
\label{sec:content_bound}
A static watermark key $sk_i = \texttt{SHA-256}(sk \,\|\, \texttt{role}_i)$ that depends only on the agent role is vulnerable to replay attacks: an adversary who intercepts a legitimately watermarked cache from query $q_1$ can re-inject it as the response to a different query $q_2$, and the verification module's check will pass because the watermark is genuinely present and the key matches.

To bind the watermark to the specific input context, we replace the static watermark key $sk_i$ with a content-bound variant $sk_i'$, derived from the per-agent key and a hash of the input query:
\begin{equation}
    sk_i' = \texttt{SHA-256}(sk_i \,\|\, \texttt{SHA-256}(q)_{1:16}),
    \label{eq:content_key}
\end{equation}
where $q$ is the input query text and the subscript $_{1:16}$ denotes the first 16 hexadecimal characters of the hash digest. All other embedding and detection equations (Eqs.~\ref{eq:dim_select}--\ref{eq:score}) are unchanged: $sk_i'$ is simply substituted for $sk_i$ wherever the key appears. The verification module independently computes the same key from the query originally sent, ensuring that a cache watermarked for query $q_1$ fails verification when checked against the key derived from $q_2$. This one-line modification reduces replay attack 
success from 100\% to 0.2\% 
(Section~\ref{sec:security}) while preserving all 
other security properties, since only the key 
derivation changes---the embedding and detection 
algorithms are identical.

\FloatBarrier

\section{Experiments}
\label{sec:experiments}

\subsection{Experimental Setup}
\label{sec:setup}

\paragraph{Model and pipeline.}
We evaluate on the LatentMAS~\cite{latentmas} framework using Qwen3-4B~\cite{qwen3} as the shared backbone across all agents.
The pipeline consists of four sequential 
agents---Planner, Critic, Refiner, and 
Judger---communicating via direct KV-cache transfer 
across $L=36$ transformer layers with $H=8$ 
key-value heads (grouped query attention with 32 
query heads) and head dimension $d=128$.

\paragraph{Datasets.}
We evaluate utility preservation on four 
benchmarks: GSM8K~\cite{gsm8k} (math reasoning), 
ARC-Challenge~\cite{arc} (science reasoning), MBPP+~\cite{mbppplus} (code 
generation), and MedQA~\cite{medqa} (medical question answering). 
Sample sizes are reported in Table~\ref{tab:accuracy}. 
Detection and security experiments use GSM8K 
($N{=}500$, yielding 1{,}500 handoff verification 
events across three inter-agent edges). All experiments 
use exact-match accuracy against gold answers.

\paragraph{Watermark configuration.}
Unless otherwise noted, we use base strengths 
$\alpha_{V,0} = \alpha_{K,0} = 0.03$, watermark 
dimensions $n_d = 16$, and detection threshold 
$\tau = 0.01$. The depth-aware schedule uses 
V-cache suppress $\gamma_V = 0.1$ at layers 0--4,
$\beta_V = 5$ at V-cache layers 24--35, and 
$\beta_K = 10$ at all K-cache layers 
(Section~\ref{sec:depth_aware}).
Content-bound keys (Eq.~\ref{eq:content_key}) are used for replay-defense experiments; all others use static watermark keys.

\paragraph{Attack suite.}
We evaluate robustness against attacks organized 
into six categories ($N{=}500$ for all experiments): 
(1)~tampering attacks: random noise scaled to 
per-layer cache magnitude at 
$\alpha \in \{0.5, 4.0\}$ for V-only and K-only, 
and $\alpha = 4.0$ for KV-both; 
(2)~the Out-of-Sight DiffMean 
attack~\cite{outofsight} on V-only 
($\alpha \in \{4.0, 8.0\}$) and 
KV-both ($\alpha \in \{0.5, 1.0, 2.0, 4.0\}$); 
(3)~forgery attacks: wrong-key verification across 
500 distinct random keys; 
(4)~deployment-relevant attacks: quantization to 
fp16 and int8; 
(5)~structural attacks: replay attacks (with 
and without content-bound keys); 
(6)~statistical invisibility: nine detection 
strategies (column-mean, variance, kurtosis, range, 
abs-mean, cross-cache correlation, KL divergence, 
head-avg variance, head-avg range) on V-cache 
(early and boost layer regions) and K-cache 
(all layers). 
Table~\ref{tab:attacks} reports detection scores and pipeline accuracy for all tested configurations.

\paragraph{Metrics.}
We report: (1)~TPR, FPR, and AUROC for 
handoff-level detection using the combined score 
$\mathcal{S}$, with individual $\mathcal{S}_V$ and 
$\mathcal{S}_K$ for channel-specific analysis; 
(2)~per-layer TPR and FPR for tamper localization; 
(3)~task accuracy for utility preservation; and 
(4)~watermark dimensions identified by a keyless 
adversary (out of $n_d = 16$; random chance 
$= 16/128 \times 16 = 2.0$) for statistical 
invisibility.
\subsection{Main Results}
\label{sec:main_results}

\paragraph{Utility Preservation.}
Table~\ref{tab:accuracy} reports task accuracy across four benchmarks spanning mathematical reasoning, science, code generation, and medical question answering. All changes fall within one sampling standard error ($\mathrm{SE} \approx 1.2\%$ at $N{=}500$, $\mathrm{SE} \approx 2.6\%$ at $N{=}300$), with three benchmarks slightly above and one slightly below the clean baseline, consistent with sampling variation.

\begin{table}[t]
\centering
\small
\caption{Task accuracy across four benchmarks (full pipeline). All changes fall within sampling error. $N{=}378$ is the full MBPP+ test set.}
\label{tab:accuracy}
\begin{tabular}{lcccl}
\toprule
\textbf{Task} & \textbf{$N$} & \textbf{Clean} & \textbf{KV-WM} & \textbf{$\Delta$} \\
\midrule
GSM8K          & 500 & 92.0\% & 92.4\% & $+$0.4\% \\
ARC-Challenge  & 500 & 92.2\% & 93.4\% & $+$1.2\% \\
MBPP+          & 378 & 67.2\% & 67.7\% & $+$0.5\% \\
MedQA          & 300 & 71.0\% & 69.0\% & $-$2.0\% \\
\bottomrule
\end{tabular}
\end{table}

\paragraph{Handoff-level Detection.}
From the 500-sample pipeline run, we extract 1{,}500 independent handoff verifications (three watermarked handoffs per sample: Planner$\to$Critic, Critic$\to$Refiner, and Refiner$\to$Judger).
Each verification independently computes $\mathcal{S}$ (Eq.~\ref{eq:score}) using the embedding agent's key $sk_i$ (correct-key score); wrong-key scores are computed with random keys for the same caches.
Table~\ref{tab:detection} summarizes detection performance.
At the pipeline level, the watermark achieves 100\% TPR with 0.0\% FPR across all 1{,}500 handoffs. The per-edge breakdown confirms consistent detection across all three handoffs ($\mathcal{S}$ ranges from 0.294 at R$\to$J to 0.314 at P$\to$C), with no single edge significantly weaker. Wrong-key verification produces scores near zero (mean $\approx 0$, max $0.0070$), while the minimum correct-key score is $0.291$, yielding a separation margin of $0.284$ and indicating complete distributional separation between authentic and forged caches.

\begin{table}[t]
\centering
\small
\caption{Handoff-level detection metrics 
(full pipeline, $N{=}1{,}500$ handoffs, $\tau{=}0.01$).}
\label{tab:detection}
\begin{tabular}{llr}
\toprule
\textbf{Scope} & \textbf{Metric} & \textbf{Value} \\
\midrule
Pipeline & TPR / FPR / AUROC & 100\% / 0.0\% / 1.0 \\
         & $\mathcal{S}$ (mean $\pm$ std) & $0.3066 \pm 0.0089$ \\
\midrule
Per-channel & $\mathcal{S}_V$ / $\mathcal{S}_K$ & $0.2055$ / $0.4077$ \\
Per-edge    & $\mathcal{S}$ (P$\to$C / C$\to$R / R$\to$J) & $0.314$ / $0.311$ / $0.294$ \\
\midrule
Wrong-key & $\mathcal{S}$ (mean / max) & $\approx 0$ / $0.0070$ \\
Correct-key & $\mathcal{S}$ (min) & $0.291$ \\
Separation  & min correct $-$ max wrong & $0.284$ \\
\bottomrule
\end{tabular}
\end{table}

\paragraph{Layer-level Tamper Localization.}
To evaluate tamper localization, we apply a V-cache steering attack (noise scaled to per-layer cache magnitude, $\alpha = 0.5$) to layers 19--35 and measure whether the per-layer V-cache correlation $c_V^{(l)}$ correctly identifies attacked layers ($N{=}500$).
Table~\ref{tab:localization} reports results. The scheme achieves perfect localization: all 17 attacked layers show statistically significant correlation drops (paired $t$-test comparing per-layer correlation scores between clean and attacked caches across $N{=}500$ samples; $t > 3.5$, $p < 0.001$), while all 19 non-attacked layers show exactly zero drop (zero variance across all $N$ samples).
The KV cache is append-only at the position level, so V-cache values at non-attacked layers are bitwise identical before and after the attack, producing exact zero drop rather than merely non-significant values. Since this attack targets only the V-cache, K-cache correlations remain unchanged at all layers ($\mathcal{S}_K = 0.433$, single-hop), confirming that the dual-channel design correctly attributes tampering to the targeted cache component. The per-layer analysis in Table~\ref{tab:localization} validates the localization capability experimentally; in deployment, a defender would compare per-layer scores against empirically calibrated baselines or use relative outlier detection across layers.

\begin{table}[t]
\centering
\small
\caption{Layer-level tamper localization 
($N{=}500$, V-cache attack on layers 19--35). Values averaged across $N$ samples.}
\label{tab:localization}
\begin{tabular}{rcccc}
\toprule
\textbf{Layer} & \textbf{No attack} & \textbf{Under attack} & \textbf{Drop} & $|t|$ \\
\midrule
0 & 0.235 & 0.235 & 0.000 & --- \\
10 & 0.177 & 0.177 & 0.000 & --- \\
18 & 0.107 & 0.107 & 0.000 & --- \\
\midrule
19$^\dagger$ & 0.136 & 0.122 & 0.015 & 21.33 \\
24$^\dagger$ & 0.363 & 0.329 & 0.034 & 56.43 \\
28$^\dagger$ & 0.297 & 0.268 & 0.030 & 45.70 \\
35$^\dagger$ & 0.146 & 0.132 & 0.014 & 19.77 \\
\midrule
\multicolumn{3}{l}{TPR (attacked)} & \multicolumn{2}{l}{100\% (17/17)} \\
\multicolumn{3}{l}{FPR (non-attacked)} & \multicolumn{2}{l}{0\% (0/19)} \\
\bottomrule
\multicolumn{5}{l}{\footnotesize $^\dagger$Attacked layer (19--35). All $|t| > 19$ ($p < 0.001$).}
\end{tabular}
\end{table}

\subsection{Security Analysis}
\label{sec:security}

\paragraph{Tamper Robustness.}
Table~\ref{tab:attacks} summarizes detection under 
blind perturbation attacks, where the adversary adds 
noise without knowing which dimensions carry the 
watermark ($N{=}500$). The KV watermark maintains 
100\% detection (1{,}500/1{,}500 handoffs) under all tested attacks. 
The dual-channel design provides mutual redundancy 
against such blind attacks: a K-only perturbation at 
$\alpha = 4.0$ degrades $\mathcal{S}_K$ from 0.408 
to 0.046 but leaves $\mathcal{S}_V$ largely intact at $0.195$; 
conversely, a V-only perturbation degrades 
$\mathcal{S}_V$ to 0.025 while $\mathcal{S}_K$ 
remains at 0.407. Even the worst case (KV-both at 
$\alpha = 4.0$) produces $\mathcal{S} = 0.036$, 
still $3.6\times$ above the detection threshold. 
A stronger adversary who identifies the watermarked 
dimensions could surgically remove the signal from 
one channel without degrading output quality; 
statistical invisibility 
(Table~\ref{tab:invisibility}) prevents this by 
ensuring that watermarked dimensions are 
indistinguishable from non-watermarked ones.

\begin{table}[t]
\centering
\small
\caption{Detection and accuracy under representative attacks 
(full pipeline, $N{=}500$, $\tau{=}0.01$). All 1{,}500/1{,}500 handoffs detected.}
\label{tab:attacks}
\begin{tabular}{llcccc}
\toprule
\textbf{Attack} & \textbf{Target} & $\mathcal{S}_V$ & $\mathcal{S}_K$ & $\mathcal{S}$ & \textbf{Acc.} \\
\midrule
No attack & --- & 0.206 & 0.408 & 0.307 & 92.4\% \\
\midrule
Steering $\alpha{=}0.5$ & V & 0.174 & 0.408 & 0.291 & 0.0\% \\
Steering $\alpha{=}4.0$ & V & 0.025 & 0.407 & 0.216 & 0.6\% \\
Steering $\alpha{=}0.5$ & K & 0.213 & 0.330 & 0.272 & 2.6\% \\
Steering $\alpha{=}4.0$ & K & 0.195 & 0.046 & 0.121 & 0.2\% \\
Steering $\alpha{=}4.0$ & KV & 0.025 & 0.047 & 0.036 & 1.0\% \\
\midrule
DiffMean $\alpha{=}4.0$ & V & 0.189 & 0.407 & 0.298 & 74.2\% \\
DiffMean $\alpha{=}8.0$ & V & 0.153 & 0.407 & 0.280 & 43.2\% \\
DiffMean $\alpha{=}4.0$ & KV & 0.182 & 0.393 & 0.287 & 1.0\% \\
\midrule
Quantization fp16 & KV & 0.206 & 0.408 & 0.307 & 93.0\% \\
Quantization int8 & KV & 0.190 & 0.401 & 0.296 & 93.0\% \\
\bottomrule
\end{tabular}
\end{table}
\paragraph{Attack Tradeoff: detection vs.\ accuracy.}
Blind steering attacks destroy output quality at all tested strengths: pipeline accuracy drops to $\leq 2.6\%$ even at $\alpha = 0.5$ (V-only: 0.0\%, K-only: 2.6\%, KV at $\alpha{=}4.0$: 1.0\%), offering no attacker tradeoff---the attack degrades both quality and detection simultaneously. Only the stronger Out-of-Sight DiffMean 
attack~\cite{outofsight}, which requires offline access to compute directional vectors, preserves accuracy at low $\alpha$. Under V-only DiffMean, the V-cache score degrades 
with increasing $\alpha$ (from $0.189$ at 
$\alpha = 4.0$ to $0.153$ at $\alpha = 8.0$, Table~\ref{tab:attacks}) but 
never drops below the detection threshold---and the 
K-cache channel remains completely unaffected 
($\mathcal{S}_K = 0.407$ at all $\alpha$), providing 
a redundant detection signal. Under the stronger 
KV-both variant, both scores degrade but detection 
remains 1{,}500/1{,}500 at $\alpha = 4.0$ 
($\mathcal{S}_V = 0.182$, $\mathcal{S}_K = 0.393$).
To quantify the attacker's dilemma, we propagate 
the KV-both DiffMean attack through the full 
four-agent pipeline and measure task accuracy 
($N{=}500$, Table~\ref{tab:diffmean_tradeoff}). 
At $\alpha = 2.0$, accuracy collapses from 
92.4\% to 1.4\% while the watermark remains 
detected ($\mathcal{S} = 0.302$); at $\alpha = 4.0$, the output degrades 
further. No operating point simultaneously evades 
detection and preserves output quality.

\begin{table}[t]
\centering
\small
\caption{DiffMean KV-both accuracy tradeoff---the only attack that preserves output quality at low $\alpha$ (full pipeline, $N{=}500$).}
\label{tab:diffmean_tradeoff}
\begin{tabular}{ccc}
\toprule
$\alpha$ & $\mathcal{S}$ & Accuracy \\
\midrule
0.0 & 0.307 & 92.4\% \\
0.5 & 0.307 & 91.8\% \\
1.0 & 0.306 & 86.8\% \\
2.0 & 0.302 & 1.4\% \\
4.0 & 0.287 & 1.0\% \\
\bottomrule
\end{tabular}
\end{table}
\paragraph{Statistical Invisibility.}
We evaluate invisibility against nine statistical 
detection strategies on both V-cache and K-cache 
($N{=}500$). Table~\ref{tab:invisibility} reports 
the number of true watermark dimensions found (out 
of $n_d = 16$) at  suppressed and boosted V-cache layers and across all K-cache layers. For each statistic, the adversary ranks all 128 dimensions by that statistic's value and selects the top 16 as suspected watermark dimensions; we report how many of these are actual watermark dimensions (chance $= 16/128 \times 16 = 2.0$). The three-tier strength schedule ensures full 
invisibility across all regions and all nine statistics: 
no statistic exceeds 3.0/16 at any layer group, 
compared to the 2.0/16 random chance baseline. 
The V-cache suppress schedule ($\gamma_V = 0.1$ at 
layers 0--4) prevents the higher-order leakage 
(variance, abs-mean) that would otherwise occur at 
early layers where natural cache magnitude is 
comparable to the perturbation. The V-cache boost ($\beta_V = 5$, lower than the K-cache's $\beta_K = 10$) keeps variance 
and abs-mean near chance at boosted layers.
The K-cache scores range from 1.0 to 2.3 across all 
nine statistics---indistinguishable from random 
chance (2.0/16)---despite the $10\times$ uniform 
boost, because K-cache natural magnitudes 
($\mathrm{std} = 1.7$--$14.1$) dwarf the boosted 
perturbation ($0.30$).
A natural extension of the keyless adversary is to average cache values across all $H$ attention heads before computing statistics, suppressing natural per-head variation. The per-head sign patterns (Eq.~\ref{eq:sign_pattern}) defeat this strategy: without the key, the watermark cancels under head-averaging ($\frac{1}{H}\sum_h \alpha \cdot \sigma^{(l,h)}_{s,j} \approx 0$). Table~\ref{tab:invisibility} confirms that head-averaged variance and range remain near chance across all layer groups.


\begin{table}[t]
\centering
\small
\caption{Statistical invisibility: watermark dimensions found by keyless adversary (chance $= 2.0/16$, $N{=}500$). V-E: V-cache early layers 0--4; V-B: V-cache boost layers 24--35; K: K-cache all layers.}
\label{tab:invisibility}
\begin{tabular}{lccc}
\toprule
\textbf{Statistic} & \textbf{V-E} & \textbf{V-B} & \textbf{K} \\
\midrule
Column-mean        & 2.0 & 2.1 & 2.0 \\
Variance           & 2.8 & 2.4 & 2.3 \\
Kurtosis           & 2.2 & 1.7 & 1.0 \\
Range              & 2.8 & 2.2 & 2.3 \\
Abs-mean           & 2.4 & 2.1 & 1.9 \\
Cross-cache corr.  & 2.4 & 2.0 & 1.8 \\
KL divergence      & 1.2 & 1.7 & 1.8 \\
Head-avg variance  & 3.0 & 2.0 & 2.2 \\
Head-avg range     & 2.6 & 1.9 & 2.0 \\
\midrule
\emph{Chance}      & \emph{2.0} & \emph{2.0} & \emph{2.0} \\
\bottomrule
\end{tabular}
\end{table}

\begin{table}[t]
\centering
\small
\caption{Cache-level indistinguishability ($N{=}500$, $\alpha{=}0.05$). A keyless adversary cannot detect watermark presence.}
\label{tab:indistinguish}
\begin{tabular}{llcc}
\toprule
\textbf{Test} & \textbf{Metric} & \textbf{Result} & \textbf{Chance} \\
\midrule
KS test & Rejection rate & 2.2\% & 5.0\% \\
Mann-Whitney U & Rejection rate & 0.0\% & 5.0\% \\
\midrule
Logistic Regression & Accuracy & $\leq$37\% & 50\% \\
MLP (128, 64) & Accuracy & $\leq$23\% & 50\% \\
\midrule
Cohen's $d$ (WM dims) & Effect size & 0.004 & $< 0.2$ \\
Cohen's $d$ (non-WM) & Effect size & 0.000 & $< 0.2$ \\
\bottomrule
\end{tabular}
\end{table}

Beyond the dimension-identification analysis above, we evaluate \emph{cache-level} indistinguishability: whether a keyless adversary can detect watermark \emph{presence} without identifying specific dimensions. Table~\ref{tab:indistinguish} summarizes three complementary tests. Two-sample Kolmogorov-Smirnov and Mann-Whitney~U tests, applied per-dimension across all 36 layers ($N{=}500$), fail to reject the null hypothesis that clean and watermarked distributions are identical (rejection rates of 2.2\% and 0.0\% at $\alpha{=}0.05$, both below the expected 5\% false positive rate). A logistic regression and two-layer MLP trained on per-dimension statistics (mean, variance, skewness, kurtosis; 514 features per layer) achieve $\leq 37\%$ and $\leq 23\%$ accuracy respectively under 5-fold cross-validation (chance $= 50\%$), confirming that even a learned adversary cannot detect watermark presence. Per-dimension Cohen's $d$ between clean and watermarked caches is negligible at the boosted layers ($d = 0.004$ for watermarked dimensions, $d \approx 0$ for non-watermarked), well below the standard $d < 0.2$ threshold for a small effect.

\paragraph{Adaptive White-box Attacks.}
Table~\ref{tab:security} summarizes results across three security scenarios. All four adaptive strategies are defeated: PCA whitening, average-and-subtract, blind dimension zeroing, and watermark overwriting all fail to evade detection. The position-dependent sign patterns ensure that dimension-level statistics do not reveal which dimensions carry the watermark, and the dual KV channels provide independent backup.

\paragraph{Forgery Resistance.}
Wrong-key detection across 500 distinct random keys produces scores statistically indistinguishable from zero (Table~\ref{tab:security}). The separation between correct-key and wrong-key scores is $0.294$, confirming that forgery without the secret key is infeasible. The wrong-key maximum here ($0.0120$) is higher than in Table~\ref{tab:detection} ($0.0070$) because this test uses shorter single-hop caches; longer pipeline caches reduce wrong-key variance.

\paragraph{Replay Defense.}

With static watermark keys, replayed caches pass verification 100\% of the time, as expected. Content-bound keys (Eq.~\ref{eq:content_key}) reduce this to 0.2\% (Table~\ref{tab:security}), while preserving legitimate verification. The residual pass is a threshold-edge case consistent with the wrong-key FPR observed in the forgery test.

\begin{table}[t]
\centering
\small
\caption{Security analysis: adaptive, forgery, and replay attacks (single-hop, $N{=}500$, $\tau{=}0.01$).}
\label{tab:security}
\begin{tabular}{llcc}
\toprule
\textbf{Category} & \textbf{Attack} & $\mathcal{S}$ & \textbf{TPR} \\
\midrule
Adaptive & PCA whitening & 0.320 & 500/500 \\
         & Avg-and-subtract & 0.317 & 20/20 \\
         & Blind dim zeroing & 0.313 & 500/500 \\
         & Overwrite 33$\times$ & 0.196 & 500/500 \\
\midrule
Forgery  & Wrong-key $\mathcal{S}$ (mean / max) & \multicolumn{2}{c}{0.0001 / 0.0120} \\
         & FPR & \multicolumn{2}{c}{2/500 (0.4\%)} \\
\midrule
Replay   & Static watermark key & --- & 500/500 \\
         & Content-bound key & --- & 1/500 \\
\bottomrule
\end{tabular}
\end{table}

\paragraph{Cascade Verification.}
\label{sec:cascade}
The KV cache is append-only, so an upstream 
agent's watermark survives at any later pipeline 
stage. This enables \emph{cascade verification}: 
the pipeline operator checks an upstream watermark 
at a downstream point. This detects a compromised 
agent that ignores its input, generates fabricated 
content, and embeds a valid watermark under its 
own key---a scenario invisible to per-handoff 
verification. Table~\ref{tab:cascade} reports results at $N{=}500$ for two compromised-agent positions. In all compromised scenarios, the upstream agents' watermarks are absent (TPR $\approx$ 0), while the compromised agent's own watermark is present (500/500). The detection rule is straightforward: if an upstream watermark is absent at a downstream point, an intermediate agent must have discarded the input. When the compromised agent re-watermarks, its own watermark identifies which agent replaced the content.

\begin{table}[t]
\centering
\small
\caption{Cascade verification (single-step generation, 4-agent cascade, $N{=}500$, $\tau{=}0.01$). ``Check for'' indicates which upstream watermark is verified at Judger's cache.}
\label{tab:cascade}
\begin{tabular}{llcc}
\toprule
\textbf{Scenario} & \textbf{Check for} & \textbf{TPR} & $\mathcal{S}$ \\
\midrule
Normal pipeline & Planner & 500/500 & 0.317 \\
\midrule
Critic ignores input & Planner & 3/500 & $\approx$0 \\
Critic re-watermarks & Planner & 3/500 & $\approx$0 \\
                     & Critic & 500/500 & 0.338 \\
\midrule
Refiner ignores input & Planner & 5/500 & $\approx$0 \\
                      & Critic & 0/500 & $\approx$0 \\
Refiner re-watermarks & Planner & 2/500 & $\approx$0 \\
                      & Critic & 0/500 & $\approx$0 \\
                      & Refiner & 500/500 & 0.315 \\
\bottomrule
\end{tabular}
\end{table}

\subsection{Ablation Studies}
\label{sec:ablation}
\begin{figure}[!t]
\centering
\includegraphics[width=\columnwidth]{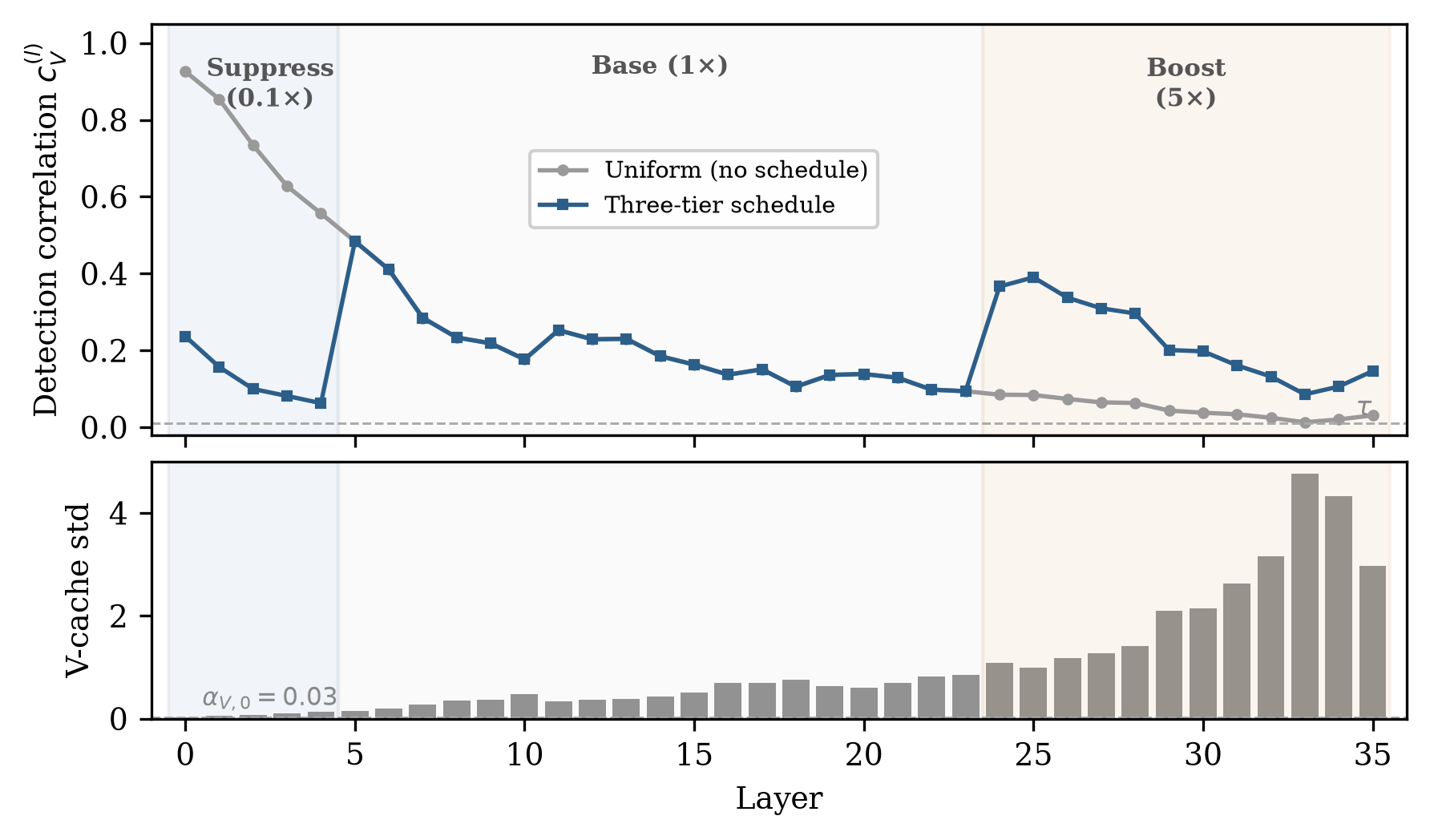}
\vspace{-8pt}
\caption{Top: V-cache detection correlation across 36 layers under uniform strength (gray, $45\times$ decay) vs.\ three-tier schedule (blue, stable above $\tau$). In the suppress region (layers 0--4), the schedule intentionally reduces correlation to ensure statistical invisibility. Bottom: V-cache magnitude (std) per layer; the dashed line shows the base perturbation $\alpha_{V,0} = 0.03$. See Appendix~\ref{sec:appendix_decay} for full per-layer data.}
\label{fig:depth_decay}
\end{figure}

\paragraph{Depth-aware Strength Schedule.}
Under truly uniform strength $\alpha_{V,0} = 0.03$, V-cache correlation decays $45\times$ across layers (Figure~\ref{fig:depth_decay}, gray line), driven by the $144\times$ growth in V-cache magnitude (bottom panel). The three-tier schedule (blue line) resolves this: suppress ($0.1\times$) at layers 0--4 reduces early-layer correlation from $0.93$ to $0.24$, preventing statistical leakage while staying above threshold; boost ($5\times$) at layers 24--35 raises late-layer correlation from $0.05$ to $0.23$, restoring localization capability. K-cache requires uniform $10\times$ boost across all layers (mean $c_K$: $0.050 \to 0.433$) due to its uniformly high magnitude (see Appendix~\ref{sec:appendix_decay} for full per-layer profiles).

\paragraph{Watermark Dimensions.}
Table~\ref{tab:ndims} reports the effect of watermark dimensionality ($N{=}100$). Increasing $n_d$ beyond 16 does not improve detection and reduces K-cache correlation, confirming that depth decay is caused by signal magnitude, not signal width. The combined score peaks at $n_d = 8$ due to higher per-dimension signal concentration, but $n_d = 16$ provides a better security-detection tradeoff with a 2.0/16 chance baseline.

\begin{table}[t]
\centering
\small
\caption{Watermark dimensions (single-hop, $N{=}100$).}
\label{tab:ndims}
\begin{tabular}{cccc}
\toprule
$n_d$ & $\mathcal{S}_V$ & $\mathcal{S}_K$ & $\mathcal{S}$ \\
\midrule
8   & 0.201 & 0.469 & 0.335 \\
16  & 0.201 & 0.433 & 0.317 \\
32  & 0.201 & 0.391 & 0.296 \\
64  & 0.202 & 0.383 & 0.292 \\
\bottomrule
\end{tabular}
\end{table}
\subsection{Comparison with Prior Work}
\label{sec:comparison}

Text-generation watermarks~\cite{kgw,threebricks} operate on token logits during decoding, and agent behavioral watermarks~\cite{agentmark,seqbehavioral,agentwm} operate on discrete action sequences---neither produces a signal in continuous KV-cache tensors, making direct quantitative comparison infeasible. The closest concurrent work is by Brito and Baquero~\cite{latentagentslie}, who study KV-cache integrity in role-sequenced multi-agent systems using HMAC-SHA256 transport manifests and magnitude quarantine. Table~\ref{tab:comparison} compares the two approaches.

\begin{table}[t]
\centering
\caption{Comparison with Brito and Baquero~\cite{latentagentslie}.}
\label{tab:comparison}
\small
\begin{tabular}{p{4.2cm}cc}
\toprule
\textbf{Property} & \textbf{\textsc{KVMark}} & \textbf{B\&B} \\
\midrule
Post-handoff substitution       & \ding{51} & \ding{51} \\
Endpoint-signed malicious KV    & \ding{51}$^\dagger$ & \ding{55} \\
Norm-matched adaptive attack    & \ding{51} & \ding{55} \\
Continuous tamper degree         & \ding{51} & \ding{55} \\
\bottomrule
\end{tabular}

\vspace{0.3em}
\parbox{\columnwidth}{\footnotesize 
$^\dagger$Via cascade verification (Section~\ref{sec:cascade}); malicious but input-preserving agents remain undetectable.\\
}
\end{table}

The HMAC approach achieves perfect binary detection 
for post-handoff substitution (774/774 honest, 
295/295 tampered) but explicitly cannot detect 
compromised endpoints that sign malicious KV 
in-band, and their magnitude quarantine is 
defeated by adaptive norm-matched attacks (0/65 
detected). Our \textsc{KVMark} addresses both scenarios: 
the keyed statistical signal is embedded in 
KV-cache values, so post-handoff substitution 
disrupts the correlation, and cascade verification 
detects endpoints that discard upstream input 
while re-embedding under their own key. The two 
mechanisms are complementary: HMAC verifies 
transport provenance, while our watermark verifies 
content integrity.

\subsection{Model Generalization}
\label{sec:generalization}
To evaluate whether the watermark generalizes beyond the primary evaluation model, we apply identical hyperparameters (no model-specific tuning) to Qwen3-8B within the same LatentMAS pipeline and benchmark (GSM8K, $N{=}500$).

\begin{table}[t]
\centering
\caption{Generalization with zero re-tuning (full pipeline).}
\label{tab:generalization}
\small
\begin{tabular}{lcccc}
\toprule
\textbf{Model} & \textbf{Clean} & \textbf{KV-WM} & $\mathcal{S}$ & \textbf{TPR} \\
\midrule
Qwen3-4B & 92.0\% & 92.4\% & 0.307 & 1500/1500 \\
Qwen3-8B & 93.4\% & 94.0\% & 0.312 & 1500/1500 \\
\bottomrule
\end{tabular}
\end{table}

Table~\ref{tab:generalization} confirms that the watermark transfers without modification: detection scores ($\mathcal{S} = 0.312$) and TPR ($1500/1500$) are statistically indistinguishable from the 4B results. The watermark operates directly on KV-cache tensors without accessing model weights, tokenizer, or architecture-specific logic; any model producing $[B, H, S, d]$ cache tensors is compatible. Cross-architecture evaluation (e.g., Llama, Mistral) requires pipeline adaptation, not watermark modification, and is left to future work.

\subsection{Computational Overhead}
\label{sec:overhead}
The watermark adds a constant perturbation to 16 
of 128 dimensions per position per layer in both 
V-cache and K-cache. Per-head sign patterns require 
generating $H$ independent sign patterns per layer 
(one per attention head) during both embedding and 
detection.
On an NVIDIA RTX~5090 GPU, 
embedding takes ${\sim}224$\,ms and detection takes 
${\sim}154$\,ms per handoff (measured at sequence length 14; both operations scale linearly with sequence length). The overhead is dominated by per-head sign pattern generation in Python and could be substantially reduced through batched GPU computation.

\section{Discussion and Future Work}
\label{sec:discussion}

\paragraph{Limitations.}

While \textsc{KVMark} establishes a robust, training-free defense for continuous Key-Value (KV) cache communication, several practical constraints highlight important directions for future research.
First, our framework currently assumes homogeneous pipeline architectures in which all participating agents share identical structural parameters (e.g., layer depth $L$, head count $H$, and head dimension $d$). In heterogeneous multi-agent systems involving diverse model families or pruned variants, direct latent handoffs require intermediate linear projections or cross-attention alignments. Adapting fragile watermarking across heterogeneous model boundaries without destroying embedded signal patterns remains an open challenge.
Second, because sign patterns are anchored to absolute token position indices, \textsc{KVMark} inherently assumes stable sequence alignment. However, operations such as dynamic KV-cache compression, sliding-window attention, or token eviction disrupt sequence indexing and trigger false-positive tamper alerts.
Third, the three-tier depth-aware strength schedule relies on parameters ($\beta_V, \beta_K$) tuned to the layer-wise magnitude growth profile of the Qwen3 architecture. While this profile generalizes seamlessly across different parameter scales (e.g., Qwen3-4B to Qwen3-8B), deploying \textsc{KVMark} on architecturally distinct model families with alternative normalization schemes (e.g., LayerNorm and RMSNorm \cite{gupta2026geometric}) requires an initial calibration step. Fortunately, this magnitude calibration is non-resource-intensive, requiring only a few forward passes on representative inputs.
Finally, fragile watermarking guarantees structural and contextual transport integrity; it inherently operates at the tensor layer rather than the semantic layer. Consequently, while \textsc{KVMark} reliably isolates in-transit tampering, man-in-the-middle modifications, and context-dropping, it cannot detect semantic or behavioral steering. A fully compromised in-band agent that reads its inputs, correctly embeds a valid watermark, but intentionally generates biased or adversarial latent content remains out of scope for transport-layer verification and requires complementary behavioral attribution mechanisms~\cite{agentmark}.

\paragraph{Future Work.}
Building on these insights, our future work includes generalizing \textsc{KVMark} to heterogeneous pipeline topologies (e.g., dynamic agent selection and branching DAGs \cite{yang2026agentnet}) and developing position-agnostic or relative-index watermarking schemes resilient to dynamic token eviction.
Furthermore, combining transport integrity verification with semantic auditing and behavioral attribution~\cite{agentmark} offers a promising solution for end-to-end security in autonomous multi-agent ecosystems.
Finally, extending fragile watermarking to alternative continuous latent communication substrates, such as compressed hidden states~\cite{interlat}, presents a compelling direction for future investigation.

\section{Conclusion}
\label{sec:conclusion}

We presented the first watermarking scheme for KV-cache integrity verification in latent 
multi-agent systems. The scheme embeds a fragile, keyed, zero-bit watermark into both V-cache and 
K-cache, enabling the verification module to verify authenticity before processing.
Comprehensive evaluation of Qwen3-4B across four benchmarks and Qwen3-8B on GSM8K demonstrates 100\% detection at $N{=}1{,}500$ handoffs, 100\% layer-level localization, no statistically significant degradation in accuracy, full statistical invisibility, and robustness to all tested attacks. 
The watermark generalizes across model scales with zero re-tuning, establishing that watermark-based integrity verification is both feasible and effective for securing latent multi-agent communication.

\bibliographystyle{plain}
\bibliography{references}

\appendix

\section{Open Science}
\label{sec:open_science}
All source code, experiment scripts, and result files are included in the supplementary material and will be made publicly available upon acceptance. The repository includes: (1)~the core watermark implementation (embedding, detection, key derivation); (2)~runnable experiment scripts reproducing every table and figure in the paper, with parser scripts that extract table values from raw logs; and (3)~original result files from the experimental runs. The experiments use publicly available models (Qwen3-4B and Qwen3-8B from HuggingFace) and datasets (GSM8K, ARC-Challenge, MBPP+, MedQA), requiring a single NVIDIA GPU with $\geq$32\,GB memory. A README maps each paper table to its corresponding script and result file.

\section{Ethical Considerations}
\label{sec:ethics}
This work proposes a defensive mechanism for verifying KV-cache integrity in multi-agent systems. The watermark is fragile by design: it detects tampering but cannot be used for persistent tracking or surveillance, as any cache modification destroys the signal. Our threat model assumes the watermarking system is operated by a trusted pipeline defender with legitimate authority over the agents, not deployed covertly. The technique does not involve human subjects, personal data, or deception. We acknowledge that watermark detection thresholds involve a precision--recall tradeoff; operators should calibrate $\tau$ to their deployment's tolerance for false positives. All experiments use publicly available models and datasets; no proprietary or restricted data was used.

\section{Depth-Dependent Signal Decay}
\label{sec:appendix_decay}

This section provides per-layer analysis motivating the three-tier strength schedule (Section~\ref{sec:depth_aware}).

\paragraph{Cache magnitude profiles.}
Table~\ref{tab:layer_magnitude} reports the standard deviation of V-cache and K-cache values across all 36 layers of Qwen3-4B ($N{=}100$). V-cache magnitude spans a $144\times$ range (0.033 at layer~0 to 4.757 at layer~33), while K-cache magnitude varies only $8\times$ (1.663 to 14.076). This asymmetry explains why V-cache requires a layer-dependent schedule while K-cache can use a uniform boost.

\begin{table}[h]
\centering
\caption{Per-layer cache magnitude (std) and detection correlation ($c_V$, $c_K$ from Eq.~\ref{eq:corr_v}--\ref{eq:corr_k}; higher = stronger detection) under uniform strength vs.\ three-tier schedule. $N{=}100$.}
\label{tab:layer_magnitude}
\small
\begin{tabular}{rcccccc}
\toprule
& \multicolumn{2}{c}{\textbf{Cache Std}} & \multicolumn{2}{c}{\textbf{Uniform}} & \multicolumn{2}{c}{\textbf{Three-Tier}} \\
\cmidrule(lr){2-3} \cmidrule(lr){4-5} \cmidrule(lr){6-7}
\textbf{Layer} & $\sigma_V$ & $\sigma_K$ & $c_V$ & $c_K$ & $c_V$ & $c_K$ \\
\midrule
0  & 0.033 & 14.08 & 0.926 & 0.069 & 0.237 & 0.542 \\
2  & 0.074 &  1.66 & 0.734 & 0.062 & 0.100 & 0.517 \\
5  & 0.149 &  4.73 & 0.484 & 0.006 & 0.484 & 0.079 \\
10 & 0.474 &  2.60 & 0.177 & 0.074 & 0.177 & 0.587 \\
15 & 0.516 &  1.95 & 0.163 & 0.043 & 0.163 & 0.393 \\
20 & 0.606 &  1.95 & 0.139 & 0.045 & 0.139 & 0.391 \\
24 & 1.087 &  1.95 & 0.085 & 0.045 & 0.366 & 0.406 \\
28 & 1.406 &  1.82 & 0.063 & 0.054 & 0.296 & 0.470 \\
32 & 3.163 &  1.85 & 0.025 & 0.044 & 0.132 & 0.384 \\
35 & 2.965 &  2.82 & 0.031 & 0.037 & 0.146 & 0.404 \\
\bottomrule
\end{tabular}
\end{table}

\paragraph{Signal decay under uniform strength.}
Under uniform perturbation $\alpha_{V,0} = 0.03$ (no boost, no suppress), V-cache correlation decays $45\times$ from layer~0 ($c_V = 0.926$) to layer~34 ($c_V = 0.020$). The decay follows directly from the signal-to-magnitude ratio: at layer~0, the perturbation ($0.03$) equals the natural V-cache standard deviation ($0.033$), producing near-perfect correlation; at layer~33, the same perturbation is $\sim\!159\times$ smaller than the cache magnitude ($4.757$), producing negligible correlation. K-cache correlation is uniformly low under uniform strength (mean $c_K = 0.050$) because K-cache magnitudes ($1.7$--$14.1$) dominate the perturbation at every layer.

\paragraph{Three-tier schedule effect.}
The three-tier schedule resolves both extremes. At early layers (0--4), suppress ($0.1\times$) reduces the perturbation from $0.03$ to $0.003$, lowering $c_V$ from $0.926$ to $0.237$---still well above threshold but no longer producing the variance and abs-mean leakage that a $0.926$ correlation implies. At late layers (24--35), boost ($5\times$) raises the perturbation from $0.03$ to $0.15$, increasing $c_V$ from ${\sim}0.05$ to ${\sim}0.23$, restoring detection capability. The K-cache uniform $10\times$ boost raises mean $c_K$ from $0.050$ to $0.433$, providing a strong independent detection channel at every layer.

\end{document}